# Impacts of Riding Comfort on the Attitudes of Riders, Drivers, and Pedestrians Toward Autonomous Shuttles

**Keke Long[1], Xiaowei Shi*[2], Zhiwei Chen[3], Yuan Wang[4], Xiaopeng Li*[1]**

[1] Department of Civil and Environmental Engineering, University of Wisconsin-Madison
[2] Department of Civil and Environmental Engineering, University of Wisconsin-Milwaukee
[3] Department of Civil, Architectural, & Environmental Engineering, Drexel University
[4] Department of Civil and Environmental Engineering, University of South Florida

**ABSTRACT**
Automated vehicle (AV) shuttles are emerging mobility technologies that have been widely piloted and deployed. Public attitude is critical to the deployment progress and the overall social benefits of automated vehicle (AV) technologies. The AV shuttle demonstration was regarded as a good way for possible attitude improvements. However, not all existing AV shuttle technologies are mature and reliable enough. Some frustrating uncomfort issues in AV shuttle demonstrations may adversely affect public attitudes of AV technologies. Studying the impact of the comfort of the demonstration on attitudinal change can help us provide guidance for future demonstrations. Thanks to the AV shuttle project piloted at Dunedin, Florida, this paper interviewed 161 people before and after taking an open road AV shuttle ride. In addition to the participant's demographic information, driving-related information and psychological-related information, we collected the AV shuttles' operation status (e.g., abrupt brake). A series of discrete outcome models were employed to estimate the factors influencing people's initial opinions before the AV shuttle ride and people's attitudes change after the ride. The most important finding is that an unsatisfactory riding experience could negatively affect participants' attitudes toward AV shuttles. Particularly, the number of abrupt brakes raises more concerns when the participants are drivers who share the rights of way with the AV shuttle compared to when the participants are pedestrians or AV shuttle riders. Moreover, the participants' attitudes toward sharing public space with AV shuttles as pedestrians and human drivers exhibited more negative changes compared with simply riding on AV shuttles. These results improve our understanding of public attitudes toward the existing AV technologies and help identify new opportunities to improve AV technology adoption.

# 1. INTRODUCTION

Automated vehicle (AV) shuttles have been made accessible to general public via a great number of pilots and demonstrations across the world in recent years. AV shuttles hold promise to sustainably improve transportation safety, mobility, and accessibility in the near future (NHTSA, 2016). Realizing these potential benefits of AV shuttles hinges on people's attitudes toward this technology. Positive attitudes will help facilitate the adoption of the new technology and realize its full potential benefits for society. Therefore, it is critical to understand people's attitudes and their relationships with contributing factors for identifying means to nurture healthy and positive attitudes toward the AV shuttle technology.

Recent studies found that people's attitudes toward AV technologies are influenced by their demographic characteristics (Ahmed et al., 2020; Krueger et al., 2016; Salonen, 2018; Sheela & Mannering, 2020; Shi et al., 2020), driving habits (Ahmed et al., 2020), and psychological characteristics (Eden et al., 2017; Liu et al., 2019; Liu & Xu, 2020; Shi et al., 2020; Xu et al., 2018). Besides these intrinsic factors that are not easy to modify, there is an increasing interest in identifying exogenous factors that are modifiable for possible attitude improvements. One such low-hanging fruit factor is the AV ride experience, which has been shown to positively influence participants' attitudes toward AVs (Kolodge et al., 2020; Pyrialakou et al., 2020; Xing et al., 2022). With these instructive findings, AV shuttle manufacturers and operators invest intensively in AV demonstrations (Dan, 2020) to shape people's attitudes and expectations to promote technology adoption via demonstrations.

Nevertheless, outcomes of real-world demonstrations were not always consistent with this optimistic expectation. In contrast, it is often shown that a non-negligible portion of participants may negatively change their attitudes after riding AV shuttles (Liu et al., 2019; Liu & Xu, 2020; Shi et al., 2020). Most studies attributed such negative attitude change to factors other than riding comfort since they simply assume all rides are homogeneous and their comfort levels are identical. However, with existing immature and unreliable AV technologies, different rides may result in heterogeneous feelings of comfort from the participants. It was reported that unexpected AV shuttle operations, such as abrupt brakes, may occasionally happen and cause riding discomfort during real-world demonstrations (Templeton, 2020). Such rides with high discomfort, as opposed to the intention of using demonstrations to encourage technology adoption, may in fact adversely affect public attitudes and discourage AV shuttle ridership. Evidently, different rides may experience different levels of comfort, which may result in heterogeneous attitude changes. For example, pleasure rides with few abrupt brakes may tend to improve participants' attitudes toward the technology after riding, while uncomfortable rides with frequent abrupt brakes may degrade their attitudes. However, as most existing studies did not recognize the heterogeneity of riding comfort across different rides, the effect of heterogeneous riding comfort on public attitude change is yet to be quantified. Without quantifying this relationship, people may mistakenly associate attitude degradation with other less relevant factors and fail to devise appropriate measures for improving public attitude and accelerating technology adoption.

Table 1 Studies involving real-world ride experience to analyze changes in public attitudes toward AV and AV shuttles.

| Study | Test ride mode | Number of participants | Demographic characteristics | Driving related information | Psychological characteristics | Comfort of the AV ride |
|---|---|---|---|---|---|---|
| Xu et al., 2018 | AV on enclosed test sites | 300 | | | √ | |
| Liu et al., 2019 | | 300 | | | √ | |
| Liu & Xu, 2020 | | 300 | | | √ | |
| Shi et al., 2020 | | 166 | √ | √ | | |
| Eden et al., 2017 | AV shuttle on public roads | 17 | | | | |
| Moták et al., 2017 | | 532 | √ | √ | | |
| Madigan et al., 2017 | | 315 | √ | √ | | |
| Salonen, 2018 | | 197 | √ | | | |
| Nordhoff et al., 2018 | | 384 | √ | | | |
| Hilgarter & Granig, 2020 | | 19 | √ | √ | | |
| This study | | 161 | √ | √ | √ | √ |

Furthermore, most existing studies focus on the attitudes of riders. But AV shuttles are expected to share the rights of way with other road users (human drivers and non-motorists such as pedestrians) in a mixed traffic environment, where the attitudes of these road users of AV shuttles also play a critical role in AV deployment. A few studies investigated pedestrians' attitudes after providing them with real-world interaction experiences with AV shuttles in field experiments and found that pedestrians' attitudes toward AV shuttles were influenced by demographic features (Deb et al., 2017; Hulse et al., 2018; Pyrialakou et al., 2020) and favorable interpretations of the AV brand (Reig et al., 2018). However, no definitive conclusions have been drawn regarding whether interactions with AV shuttles lead to the increased trust of AV shuttles from pedestrians. One exemplary program is the three-year BikePGH program in Swiss. While several studies showed that interaction experience improved pedestrians' attitudes towards AV shuttles in the BikePGH program (Das et al., 2020; Penmetsa et al., 2019; Rahman & Dey, 2022), other studies did not report substantial improvements (Xing et al., 2022). More studies are thus needed to reconcile these contradictory results. Regarding human drivers' attitudes of AVs, studies based on real-world interaction experiences were not found by the research team.

The above discussions identify two critical research gaps in the literature, which are important prerequisites for deriving a further understanding of and developing effective strategies to promote public acceptance of AV shuttles. First, existing studies have not fully considered the influence of the comfort of AV ride experience on public attitudes toward AV shuttles. Second, to our best knowledge, no study provides a holistic investigation of the public attitudes toward AV shuttles, including AV riders, human drivers, and other road users such as pedestrians who share the rights of way with AV shuttles.

To fill these gaps, this study used the AV shuttles operating in Dunedin, FL, as a field experiment to investigate public attitudes toward AV shuttles, including riders, pedestrians, and human drivers (we call them drivers in the remainder of this paper for the convenience of the illustration). A survey was carried out to measure the participants' attitudes of the AV shuttle before and after riding on the AV shuttle. Different from existing studies, the survey collected information from the participants from three perspectives, asking them to answer the same set of questions as riders, pedestrians, and human drivers before and after a ride. Apart from the information that was commonly collected in existing studies (such as individual characteristics, and attitudes on riding comfort before and after rides), the survey collected data related to the comfort of each participant's ride. The data allow the investigation of exogenous factors influencing the participants' initial attitude and attitude change towards AV shuttles with statistical models. The initial attitude was modeled to offer a baseline for understanding the participants' attitudes before the field experiment. The attitude change was to unveil both intrinsic and exogenous influencing factors that change public attitudes toward AV shuttles. To this end, a set of likelihood ratio tests was conducted, revealing statistically different differences in participants' initial attitude and attitude change toward AV between the three perspectives. That is, participants showed different initial attitudes and attitude changes toward AV shuttles when answering the questions in the survey as riders, pedestrians, and drivers. Thus, models were estimated separately from the perspectives of riders, pedestrians, and drivers. Specifically, ordered probit models with random parameters are adopted to model the initial attitude toward AV shuttles, and multinomial logit models are employed to explain the attitude change.

With these, this study makes two contributions to the literature. First, it studies how the comfort of the ride experience affects public attitudes toward AV shuttles. Results show that an unsatisfactory riding experience could negatively affect public attitudes toward AV shuttles. The unsatisfactory experiences were mostly caused by existing immature AV technologies (abrupt brakes, driving slowly, being too conservative at intersections). These results, thus, offer evidence to our argument that existing AV technologies are too immature to realize the expected benefits of AV shuttle demonstrations. AV pilot program planners need to be aware of the necessity of test ride comfort to minimize the possible negative outcomes (such as causing distrust) of their programs. Second, this study investigates people's willingness to share the rights of way with AV shuttles as pedestrians and drivers. The analysis revealed that the participants' attitudes toward sharing the rights of way with AV shuttles as pedestrians and human drivers exhibited more negative changes than simply riding on AV shuttles. This result indicates that obtaining trust from pedestrians and human drivers to share public roads with AV shuttles is probably a more critical challenge in the wide deployment of AV shuttles in the real world. AV operators might consider pedestrians' and drivers' concerns when promoting AV shuttles by reducing distrust and uncertainty when pedestrians and drivers interact with AV shuttles. Overall, these results depict a complete picture of the public acceptance of existing AV technology. They offer important insights that AV pilot program planners and managers can draw on when planning future AV pilot programs. They are also useful for practitioners to design future AV deployment strategies in the real world.

The rest of this paper is organized as follows. Section 2 and Section 3 introduce the data collection and modeling approaches. Section 4 discusses the model estimation results for initial public attitudes and attitude change toward AV shuttle comfort. Section 5 concludes the work and points out future research directions.

## 2. DATA COLLECTION

The primary goal of this study is to unveil intrinsic and exogenous factors influencing the public attitude toward AV shuttles and attitude change. To this end, we first conducted a survey collecting basic information and people's attitude toward the AV shuttle before and after the ride. This section briefly introduces the data collection process.

The attitude data used in this study are from the survey data collected in Dunedin from May to July 2022. Figure 1 shows the AV shuttle operated by Pinellas Suncoast Transit Authority (PSTA). The AV shuttles are electric vehicles with a capacity of 10 passengers (8 seats and 2 extended seats). There are no steering wheel or pedals in AV shuttles. The AV shuttles will operate following a pre-determined route and react to the surrounding traffic environment in real-time. A safe guard is always on board to turn the vehicle on and off and monitor the operation of the AV shuttle. The maximum speed of the AV shuttles is 20 km/h. Figure 2 shows the route of the AV shuttles. The circle route is around 1,600 meters, with eight stations along a two-lane road.

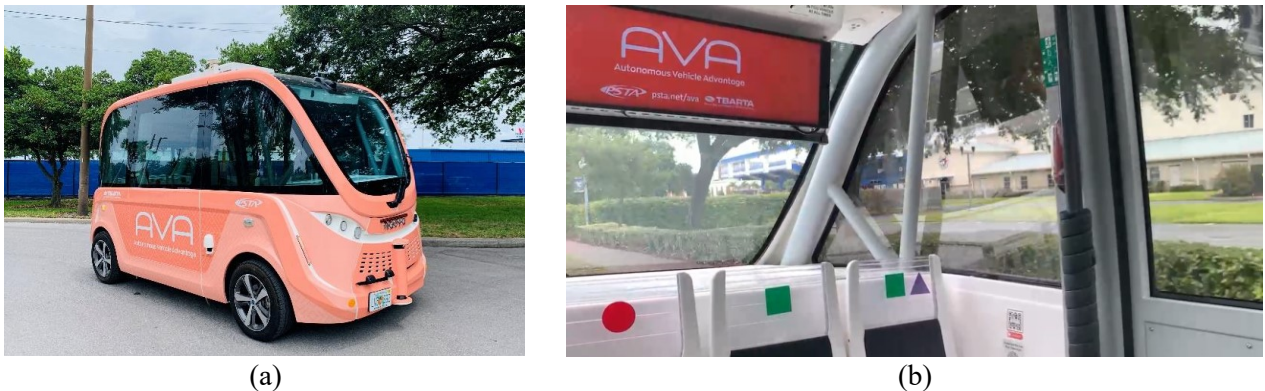

(a) (b)

Figure 1 AV shuttle operated by PSTA (a. Exterior look. b. Interior view while driving.)

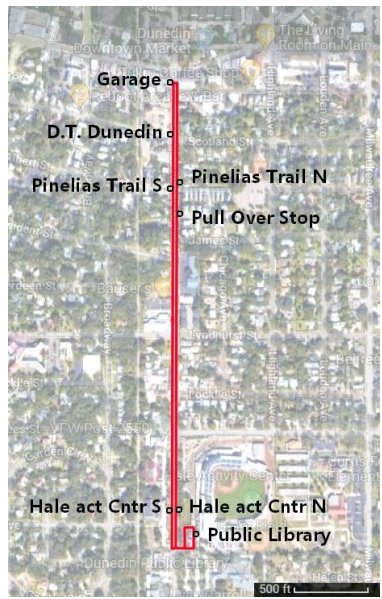

Figure 2 Route of the AV shuttle (Source: Google maps)

A questionnaire was designed to collect public attitudes of AV shuttles (see Appendix for details). The questionnaire has three parts: 1) basic information about the participants (including demographics of the participants (Q1-Q4), daily commute information (Q5, Q6), and previous experience with emerging vehicle technologies (Q7, Q8)), 2) attitudes toward the safety of the AV technology before taking the AV ride (Q9-Q11), 3) attitudes toward the comfort of the AV technology after taking the AV ride (Q12-Q14). In the second and third parts, participants were provided a 5-point scale ranging from 'very uncomfortable' to 'very comfortable' to show their feeling of comfort. There were three perspectives of attitude. The first perspective asked participants whether they felt comfortable as AV shuttle riders. The second perspective asked participants whether they felt comfortable as pedestrians walking on AV shuttle-passed roads. The third perspective asked participants whether they felt comfortable as drivers sharing the right of way with AV shuttles.

All participants will be informed that the vehicle was a self-driving AV shuttle before answering the questionnaires. The first and second parts are filled before participants get on the AV shuttle or during the first few minutes of the ride. Participants filled out the third part of the questionnaire after riding in the AV shuttle for at least 10 minutes. Both paper-based and web-based questionnaires with the same questions were used.

Additionally, during the survey, information related to the ride experience (including weather, test time, test date, and the number of emergency braking) was manually recorded by a well-trained graduate research assistant who stayed in the AV shuttle.

As mentioned in the introduction, the questionnaire answering data has the potential to capture the influencing factors of participants' attitudes before the field experiment. The information related to the ride experience has the potential to unveil previously overlooked influencing factors that change public attitudes toward AV shuttles.

## 3. METHODS

This section presents the statistical evaluation criteria, and methodological approaches for initial attitudes and for attitude change to investigate factors influencing the public attitude toward AV shuttles and attitude change.

### 3.1. Statistical evaluation criteria

Since our questionnaire involved participants' attitudes before and after the ride experience and attitudes from three perspectives (riders, pedestrians, and drivers), we first had to confirm whether different models are necessary for those different attitudes. This study conducted a likelihood ratio test (Washington et al., 2020) to assess whether statistically significant differences exist between riders, pedestrians, and drivers in terms of their initial attributes and attribute changes toward AV ride comfort. The likelihood ratio test is a statistical test of the goodness of fit between models. Separate models of two subgroups are compared to the model of the whole group to see if the two separate models of two subgroups fit the dataset significantly better. If so, there are significant differences between the two subgroups, and different models are necessary. Estimation results from three models are used: a model for all participants, a model for group A (e.g., the attitude before the test ride), and a model for group B (e.g., the attitude after the test ride). With these model estimates, a likelihood ratio test was conducted as follows:

$$\chi^2 = -2[LL(\beta)_{all} - LL(\beta)_A - LL(\beta)_B]. \qquad (1)$$

where $LL(\beta)_{all}$ is the log-likelihood at the convergence of a model using the data of all participants, $LL(\beta)_A$ and $LL(\beta)_B$ are the log-likelihood at the convergence of models using the data of group A and group B. The resulting $\chi^2$ is $\chi^2$ distributed, with degrees of freedom equaling to the difference between the numbers of parameters in the models. Chi-Square value is applied to determine whether the null hypothesis stating that the parameters in two models are equal can be rejected (Chen & Li, 2021).

### 3.2. Methodological approach for initial attitudes

The participants' initial attitude of the comfort of the AV shuttles is studied first. Given the ordered nature of the available responses to this question, an ordered probability modeling approach is adopted for the analysis.

Traditional ordered probability models are specified by defining an unobserved variable, $z$, for each observation $n$ as the linear function,

$$z_n = \beta X_n + \varepsilon_n \tag{2}$$

where $X_n$ is a vector of explanatory variables determining the discrete answers for participant $i$, $\beta$ is the vector of estimable parameters, $\varepsilon_n$ is the disturbance term. Note that only one participant chooses "very uncomfortable". To ensure an adequate sample size, the two negative altitude choices ("very uncomfortable" and "uncomfortable") options are summarized. The non-numerical ordered initial attitudes, $y_n^B$ (B represents before the test ride), are converted to integers without loss of generality (i.e., 1 = "very uncomfortable" and "uncomfortable", 2 = "neutral", 3 = "comfortable", 4 = "very uncomfortable").

$$\begin{cases} y_n^B = 1, z_n < \mu_0 \\ y_n^B = 2, \mu_0 < z_n < \mu_1 \\ y_n^B = 3, \mu_1 < z_n < \mu_2 \\ y_n^B = 4, \mu_2 < z_n \end{cases} \tag{3}$$

where $\mu_0$ to $\mu_2$ are estimable thresholds that define $y_n^B$. They are estimated jointly with the model parameters $\beta$. With this, if $\varepsilon_n$ is assumed to be normally distributed across observations with a mean equal to 0 and variance equal to 1, the ordered selection probability is

$$\begin{cases} P(y_n^B = 1) = \Phi(-\beta X) \\ P(y_n^B = 2) = \Phi(\mu_1 - \beta X) - \Phi(-\beta X) \\ P(y_n^B = 3) = \Phi(\mu_2 - \beta X) - \Phi(\mu_1 - \beta X) \\ P(y_n^B = 4) = 1 - \Phi(\mu_2 - \beta X) \end{cases} \tag{4}$$

where $\Phi()$ is the standardized cumulative normal distribution.

To capture the unobserved heterogeneity of explanatory variables, a random parameters approach (Washington et al., 2020) is integrated into the model. Thus, the parameters can vary across observations. This will help avoid potential bias and erroneous statistical inferences. The estimable parameters are written as,

$$\beta_n = \beta + w_n \tag{5}$$

where $\beta_n$ is a vector of estimable parameters that potentially varies across observations $n$, $\beta$ is the vector of mean parameter estimates across all observations, and $w_n$ is a vector of randomly distributed terms. Estimation of the random parameters ordered probit is undertaken by the simulated maximum likelihood approach.

### 3.3. Methodological approach for attitude change

Although previous research found that the initial attitudes might cause the anchoring effects that will affect the participants' final attitudes (Sheela & Mannering, 2020), the likelihood ratio test verifies that participants' attitudes were unstable between their initial assessment of AV safety and their final assessment. With this result, a series of models are estimated to understand which factors determine the likelihood of participants shifting from their initial attitudes about AV safety. The ordered probit estimation results provide insights into the factors influencing initial attitudes on AV safety. Here we focus on attitude change from different initial attitudes.

After excluding the incomplete questionnaire answers without feedback for specific questions (e.g., people without a driver's license could not perceive the comfort level of being a driver), there were a total of 25 kinds of attitude shifting. However, due to the limited number of observations, the number of several attitude shifting was too small. For example, among the attitudes from the perspective of pedestrians, only three people changed from 'uncomfortable' to 'normal', and only one changed from 'uncomfortable' to very 'uncomfortable'. Continuously using the statistical models to estimate the influence factors for these two kinds of attitude change may lose significance. Therefore, only the attitude change is studied in the rest of this paper. Three kinds of attitude changes are studied: negative change, non-change, and positive change. Among three kinds of attitudes, if the post-ride attitude changes to a more negative attitude, the post-ride option is categorized as negative; if the attitude does not change, the post-ride option is categorized as no-change; if the post-ride attitude changes to a more positive attitude, the post-ride option is categorized as a positive change.

We applied the multinomial logit model for attitude change. The probability of attitude change is:

$$P(0) = \frac{e^{V_0}}{e^{V_0}+e^{V_1}+e^{V_2}} \tag{6}$$

$$P(1) = \frac{e^{V_1}}{e^{V_0}+e^{V_1}+e^{V_2}} \tag{7}$$

$$P(2) = \frac{e^{V_2}}{e^{V_0}+e^{V_1}+e^{V_2}} \tag{8}$$

where $P(0)$, $P(1)$ and $P(2)$ are the probabilities that participants have a negative change, non-change, and positive change after the test ride. $V_0$, $V_1$ and $V_2$ are corresponding indirect utility functions.

## 4. RESULTS AND DISCUSSION

This section discusses the results of statistical models. Section 4.1 presents general descriptive statistics of the survey data. Section 4.2 presents results from the likelihood ratio test, which shows the necessity of separately modeling riders, pedestrians, and drivers and pre-ride and post-ride attitudes. Sections 4.3 and 4.4 show the evaluation results of initial attitudes and attitude change, respectively.

### 4.1. Descriptive statistics

Data from 161 participants were obtained. The age and gender distributions of the participants are generally consistent with those from the local population obtained from the American Community Survey 5-Year Data (shown in Figure 3), which shows the participants were representative of the general population. To ensure accuracy, teenagers under ten did not participate in the questionnaire.

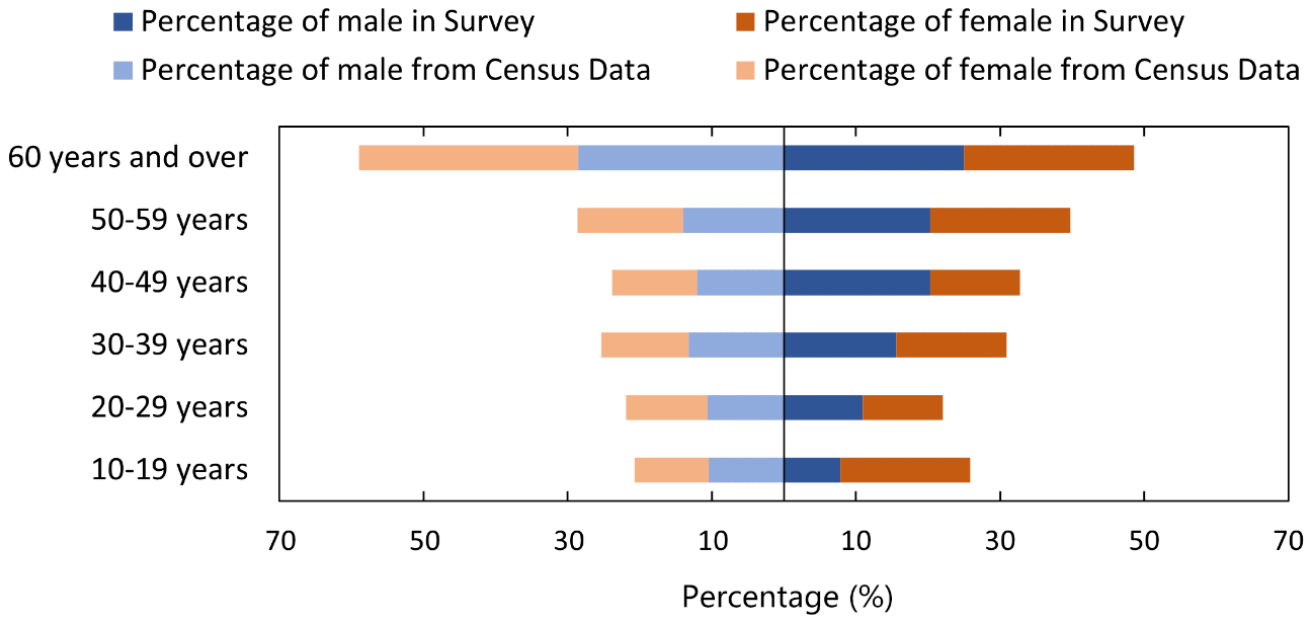

Figure 3 Comparison of gender and age distribution of participants and the local population

Participants have different attitudes from the three perspectives. Figure 4 illustrates people's initial attitude of comfort from three different roles. The percentage of participants who choose 'Very uncomfortable' or 'uncomfortable' as pedestrians and drivers exceeds the percentage of participants who choose 'Very uncomfortable' or 'uncomfortable' as AV shuttle riders. This may be because pedestrians and drivers are more vulnerable to a crash with AV shuttles.

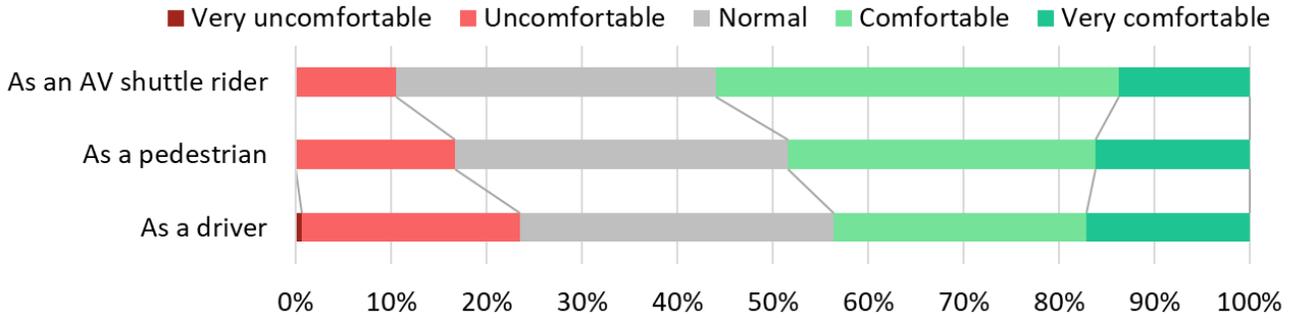
Figure 4 People's initial attitude of comfort from three different roles.

Figure 5 shows people's attitudes of comfort from three different roles after test rides. The percentage of participants who choose 'Very uncomfortable' or 'uncomfortable' as AV shuttle riders and drivers exceeds the percentage of participants who choose 'Very uncomfortable' or 'uncomfortable' as pedestrians. This may be because AV shuttles show sensitive collision avoidance behavior in operation. This collision avoidance behavior enhances the comfort of pedestrians. However, too-sensitive collision avoidance causes abrupt brakes. The sharp braking can make the AV shuttle riders feel uncomfortable and increase the danger of rear-end collision to vehicles following the AV shuttle.

Both the attitude before and after the test ride show people are more conservative about driving near AV shuttles rather than walking around AV shuttles. However, previous survey-based studies had different conclusions about the safety perception of walking near an AV compared to driving near an AV (Pyrialakou et al., 2020). One possible reason for this discrepancy is that real AV ride experience is necessary. When the observers saw the real AV, the attitude answers would naturally receive the effect of the slow driving speed of the AV. Future research could also further investigate what AV shuttles' characteristics cause people to feel uncomfortable from pedestrians' and drivers' perspectives.

The above qualitative comparisons of people's attitudes of comfort show a clear difference among the three perspectives. In the next sections, the detailed factors of the attitude of comfort are quantitatively analyzed in separate statistic models of the three perspectives.

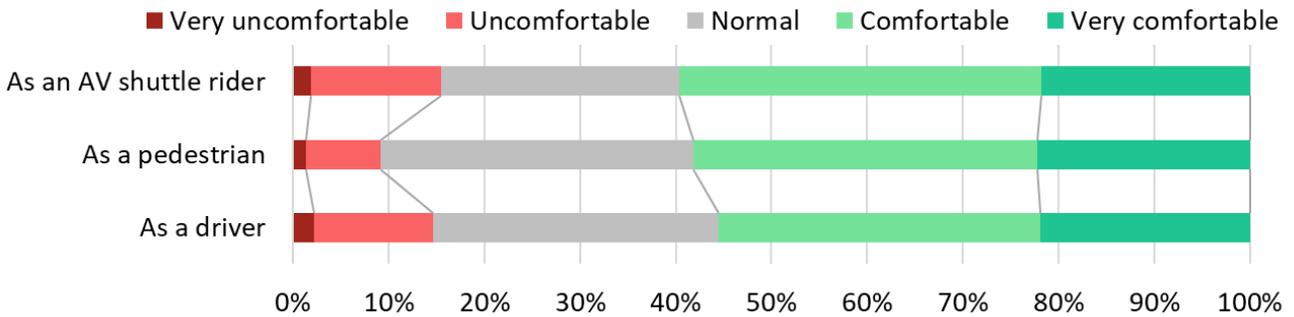
Figure 5 People's attitude of comfort from three different roles after test rides.

## 4.2. Likelihood ratio test result

The likelihood ratio test allows us to determine whether different models are needed for attitudes before and after the test ride and the attitudes from three perspectives. The likelihood ratio test results are summarized in Table 2. Two separate attitudes, Attitude 1 and Attitude 2, are evaluated to see if different statistical models are required. Comparing the riders' pre-ride attitudes with the pedestrians' pre-ride attitudes, we found a $\chi^2$ statistic of 9.740 with 4 degrees of freedom. This indicates that we are more than 95.5% confident to reject the hypothesis that riders' pre-ride attitudes model and pedestrians' pre-ride attitude model are the same. Similarly, comparing the riders' pre-ride attitudes with riders' post-ride attitudes gives a $\chi^2$ statistic of 23.868 with 6 degrees of freedom. This suggests we are more than 98.7% confident to reject the hypothesis that riders' pre-ride attitudes model and pedestrians' pre-ride attitude model are the same. These results provide substantial evidence that the riders' pre-ride attitude model is not the same as the post-ride attitude model. And the pre-ride attitude is not stable from the perspective of riders, pedestrians, and drivers. Based on these results, different models could be adopted for pre-ride attitude and post-ride attitude, and also different models for different perspectives.

Table 2 Likelihood ratio test results between different attitudes ($\chi^2$ values with degrees of freedom in parenthesis and confidence level in brackets)

| Attitudes 1 | Attitudes 2 | | |
|---|---|---|---|
| | Riders' pre-ride attitudes | Pedestrians' pre-ride attitudes | Drivers' pre-ride attitudes |
| Riders' pre-ride attitudes | - | 9.740 (4) [>95.5%] | 24.868 (6) [>99.9%] |
| Pedestrians' pre-ride attitudes | 9.740 (4) [>95.5%] | - | 12.825 (5) [>97.5%] |
| Drivers' pre-ride attitudes | 24.868 (6) [>99.9%] | 12.825 (5) [>97.5%] | - |
| Riders' post-ride attitudes | 23.868 (6) [>98.7%] | - | - |
| Pedestrians' after-ride attitudes | - | 10.740 (4) [>96.5%] | - |
| Drivers' post-ride attitudes | - | - | 12.225 (5) [>97.5%] |

### 4.3. Model estimation results of initial attitudes

The estimation results of the ordered probit model for the initial attitude of comfort in AV shuttles and the corresponding marginal effects are shown in Table 3. Five variables were found to significantly influence initial attitudes toward AV shuttles.

Participants over the age of 60 are more likely to have a negative attitude of the AV shuttle. The marginal effect shows that participants over the age of 60 are 7.1% less likely to have a "Very comfortable" attitude than younger people. This result matches those from earlier studies (Madigan et al., 2017). However, Nordhoff et al. (2018) pointed out that seniors have a higher intention to use AV shuttles. This disagreement may come from other hidden factors, including the route of the AV shuttle, the schedule, etc., and deserves further study.

Participants with high income (annual personal income higher than 100k) are inclined to have a negative attitude of the AV shuttle. Participants with high incomes have a 6.4% higher probability of having negative attitudes than others with lower incomes. Numerous studies have shown that respondents' economics influence their attitudes toward using AV shuttles, but their findings are completely opposite. Previous research found that people with high incomes are more negative toward automated techniques (Shi et al., 2020) and are less concerned about AV crashes and equipment/system failure (Ahmed et al., 2020). A cross-country survey also found that respondents from countries with higher GDP per capita were more critical of using AV shuttles. However, other research indicates that wealthy groups are more likely to adopt AVs (Sheela & Mannering, 2020) or AV shuttles (Cunningham et al., 2019). Determining the impact of personal wealth on attitudes will help target users and set ticket prices.

Participants with long commute times (commute time is longer than 30 minutes) are inclined to have a negative attitude of the AV shuttle. Participants with long commute times have a 22.4% higher probability of having negative attitudes than shorter commute times. This factor was not included in previous studies. A previous study found that the higher a person's monthly fuel cost (Shi et al., 2020), the more negative the attitude toward AV. Although there is no strict relationship between commuting time and fuel cost, these similar findings suggest that AV may be more suitable to be accepted as a short-distance travel mode in the future.

Participants who drive alone in commuting and are highly interested in new technologies are more apparent to have a positive attitude of the AV shuttle. There are two possible reasons for this result. First, driving alone is boring, and easy to get distracted, while AV technology can avoid the safety risks caused by driver distraction. Second, people who drive by themselves are very familiar with driving, thus they have confidence in AV technology. People who do not drive alone (share ride, transit, bicycle) lack driving experience and therefore do not understand and trust new technologies about driving.

Participants who are highly interested in new technologies have a 9.6% higher probability of having a "Very comfortable" attitude than people less interested in new technologies.

Two of the five variables were found to produce normally distributed random parameters with statistically significant standard deviations indicating significant unobserved heterogeneity among the participants. The long commute time indicator (1 if the participant's commute time is longer than 30 minutes, 0 otherwise) has an estimated mean parameter of -1.137 and a standard deviation of 0.988. This demonstrates that the effect of this variable increases the likelihood of having a 'normal' attitude by roughly 55.96%. This heterogeneity may stem from different expectations of the AV shuttle's effectiveness in fastening commuting. Part of the participants with long commute times regards AV shuttles as a complementary traffic mode of the current transportation system that can quickly transport people from homes to stations. Other groups of people may not see AV shuttles as comfortable as they do not compare favorably with existing buses. Therefore, the attitudes of people with long commutes toward AV shuttle are worth studying in depth to investigate whether AV shuttle is likely to be accepted as a new solution to long commutes. The other random parameter is the highly interested in new technologies indicator (1 if the participant is highly interested in new technologies, 0 otherwise), it is found with a mean of 0.564 and a standard deviation of 0.632, indicating that the effect of this variable increased the likelihood of having a 'normal' attitude for roughly 11.70%. This indicates that even among people who are willing to try new technologies, the attitude of AV is not the same.

Table 3 Ordered probit model for people's initial attitudes toward comfort as riders.

| Variable Description | Mean | Estimated Parameter | t Statistic | Marginal Effects | | | |
|---|---|---|---|---|---|---|---|
| | | | | Very uncomfortable/ Uncomfortable | Normal | Comfortable | Very comfortable |
| Constant (Standard deviation of parameter distribution) | | 1.503 (0.169) | 6.84 (1.88) | | | | |
| Senior indicator (1 if the participant is over 60-year-old, 0 otherwise) | 0.255 | -0.503 | -2.36 | 0.071 | 0.126 | -0.125 | -0.071 |
| High-income indicator (1 if the participant has an annual personal income higher than 100k, 0 otherwise) | 0.292 | -0.483 | -2.26 | 0.064 | 0.126 | -0.122 | -0.068 |
| Long commute time indicator (1 if participant's commute time is longer than 30 minutes, 0 otherwise) (Standard deviation of parameter distribution) | 0.161 | -1.137 (0.988) | -4.27 (3.80) | 0.224 | 0.196 | -0.308 | -0.112 |
| Drive alone indicator (1 if the main commute mode of the participant is driving alone, 0 otherwise) | 0.503 | 0.465 | 2.40 | -0.059 | -0.128 | 0.116 | 0.072 |
| Highly interested in new technologies indicator (1 if the participant is highly interested in new technologies, 0 otherwise) (Standard deviation of parameter distribution) | 0.435 | 0.564 (0.632) | 2.97 (4.36) | -0.061 | -0.158 | 0.122 | 0.096 |
| Threshold 1 | | 1.446 | 7.53 | | | | |
| Threshold 2 | | 2.939 | 12.30 | | | | |
| Number of observations | | | | 161 | | | |
| Log-likelihood at zero [$LL(0)$] | | | | -199.607 | | | |
| Log-likelihood at convergence [$LL(\beta)$] | | | | -184.200 | | | |
| McFadden $\rho^2 [1 - LL(0)/LL(\beta)]$ | | | | 0.077 | | | |

The estimation results of the ordered probit model for people's initial attitude of comfort as pedestrians are shown in Table 4. Similar to the modeling result of people's initial attitude of comfort in AV shuttle, short commute time and being highly interested in new technologies are still significant characteristics that drive people to have initial positive attitudes as pedestrians.

The results also suggest that the participants with long commute times, automated riding experience, and high interest in new technologies had considerable unobserved heterogeneity in their initial attitudes toward AV shuttles. Two reasons might explain why automated riding experience has considerable unobserved heterogeneity in initial attitudes toward AV shuttles, especially in our sample. The first reason is the various expectations among participants. If their previous automated riding experience met their expectations, their attitude of comfort towards AV shuttles would be positive. If the previous experience did not meet expectations, it is vice versa. Further investigation of how people expected the AV shuttle would be needed to verify this conjecture. Another possibility is that participants had previously experienced automated driving at different technical levels. We did not set a strict definition of automated riding experience, which means those participants who had only experienced vehicles with low automation levels that are currently commercially available (e.g., automatic cruise, lane keeping function) are regarded as having automated riding experience. Thus, the AV shuttle, as an L3 level of automation technology, is regarded more advanced than the vehicles with low automation levels and more comfortable. Other participants may have experienced or know about better autonomous driving (e.g., commercial self-driving vehicles such as Tesla that can plan their paths). Therefore, the AV shuttle, which follows a fixed track, seems less comfortable. This result indicates that the comfort of the AV shuttle riding experience has an important impact on the public and deserves in-depth analysis. If those currently available AV technologies cannot meet the user expectations, this may leave people with a negative initial impression of other AV technologies in the follow-up process.

Table 4 Ordered probit model for people's initial attitudes toward comfort as pedestrians.

| Variable Description | Mean | Estimated Parameter | $t$ Statistic | Marginal Effects | | | |
|---|---|---|---|---|---|---|---|
| | | | | Very uncomfortable/ Uncomfortable | Normal | Comfortable | Very comfortable |
| Constant (Standard deviation of parameter distribution) | | 1.018 (0.469) | 6.02 (4.85) | | | | |
| Long commute time indicator (1 if participant's commute time is longer than 30 minutes, 0 otherwise) (Standard deviation of parameter distribution) | 0.168 | -0.924 (0.527) | -3.42 (2.10) | 0.203 | 0.123 | -0.234 | -0.092 |
| Automated riding experience indicator (1 if the participant has automated riding experience before test ride, 0 otherwise) (Standard deviation of parameter distribution) | 0.768 | 0.536 (0.411) | 2.32 (1.95) | -0.142 | 0.111 | 0.103 | -0.072 |
| Highly interested in new technologies indicator (1 if the participant is highly interested in new technologies, 0 otherwise) (Standard deviation of parameter distribution) | 0.445 | 0.821 (0.908) | 4.28 (5.63) | -0.127 | -0.192 | 0.182 | 0.137 |
| Threshold 1 | | 1.382 | 8.48 | | | | |
| Threshold 2 | | 2.731 | 12.53 | | | | |
| Number of observations | | | | 155 | | | |
| Log-likelihood at zero [$LL(0)$] | | | | -205.542 | | | |
| Log-likelihood at convergence [$LL(\beta)$] | | | | -191.137 | | | |
| McFadden $\rho^2 [1 - LL(0)/LL(\beta)]$ | | | | 0.070 | | | |

The estimation results of the ordered probit model for people's initial attitude of comfort as drivers are shown in Table 5. People with short commute times and who are highly interested in new technologies always have a more positive attitude toward AV shuttles than others. Note that the commute time indicator is statistically significant in all three initial models, indicating that people with long commute times are less likely to embrace automated driving technology in the future. This may be because of the low operating speed of AV shuttles.

Females and participants over the age of 60 have a more negative attitude of comfort as drivers. Previous studies also reported more conservative attitudes among females and seniors (Hulse et al., 2018). These attitudes might be affected by the respondent's confidence in their own driving ability, as recent studies suggest that females are less likely than males to be confident drivers (Bergdahl, 2005).

The statistically significant random parameters were only found in the senior indicator. This implies that attitudes toward AV shuttles vary greatly among seniors. Some seniors who can drive themselves may view AV shuttles as less comfortable and convenient than private cars, while some seniors who cannot drive themselves may welcome AV shuttles as their travel opportunities increase.

Table 5 Ordered probit model for people's initial attitudes toward comfort as drivers.

| Variable Description | Mean | Estimated Parameter | $t$ Statistic | Marginal Effects | | | |
|---|---|---|---|---|---|---|---|
| | | | | Very uncomfortable/ Uncomfortable | Normal | Comfortable | Very comfortable |
| Constant | | 1.061 | 4.65 | | | | |
| Senior indicator (1 if the participant is over 60-year-old, 0 otherwise) (Standard deviation of parameter distribution) | 0.273 | -0.523 (0.778) | -2.36 (3.59) | 0.147 | 0.032 | -0.080 | -0.099 |
| Female indicator (1 if the participant is female, 0 otherwise) | 0.561 | -0.344 | -1.73 | 0.094 | 0.036 | -0.051 | -0.080 |
| Long commute time indicator (1 if participant's commute time is longer than 30 minutes, 0 otherwise) | 0.165 | -0.818 | -2.94 | 0.254 | 0.015 | -0.134 | -0.134 |
| Highly interested in new technologies indicator (1 if the participant is highly interested in new technologies, 0 otherwise) | 0.453 | 0.567 | 2.92 | -0.158 | -0.060 | 0.083 | 0.135 |
| Threshold 1 | | 1.040 | 7.59 | | | | |
| Threshold 2 | | 1.941 | 10.12 | | | | |
| Number of observations | | | | 139 | | | |
| Log-likelihood at zero $[LL(0)]$ | | | | -188.994 | | | |
| Log-likelihood at convergence $[LL(\beta)]$ | | | | -176.838 | | | |
| McFadden $\rho^2 [1 - LL(0)/LL(\beta)]$ | | | | 0.064 | | | |

Comparing people's initial attitudes from three perspectives, the results suggested that the influencing factors differed across three perspectives. Some factors were found to be only significant in particular perspectives.

Females showed a significantly negative attitude towards AV shuttles as drivers. However, in the attitudes of riders and pedestrians, gender did not significantly affect attitudes. This difference may be caused by women's lack of confidence in their driving skills rather than their attitude toward AV shuttles. The comfort attitudes might be affected by the respondent's confidence in their driving ability. Females are less likely than males to be confident drivers reported by a previous study (Bergdahl, 2005).

The automated riding experience only made participants feel more comfortable from the passenger's perspective but did not affect rider or driver attitudes, which suggests that test rides on AV shuttles can largely dispel pedestrian anxiety. This may be because people are the most vulnerable road users. Therefore, people feel the least comfort from the pedestrian's perspective when they have no experience with AV shuttles. This suggests the importance of focusing on pedestrian-oriented publicity to dispel pedestrians' worries.

Two factors only affected riding the AV shuttle, but not the perspective of sharing the road with the AV shuttle. This may be determined by the traffic mode of the shuttle. People with high incomes do not want to ride in AV shuttles. This may be because high-income people, who can afford private cars, think that shuttles are not as comfortable as driving and thus do not want to take shuttles, but there is no negative attitude towards AV technology. Therefore, people with high incomes are not repulsed by AV technology. They are just not potential users of shuttles. Future research could explore their acceptance of other forms of AV, such as AV taxis and private AV shuttles. The other factor, driving alone, only positively impacts the attitude of AV shuttle riders. This is possible because people who often drive alone may regard driving as a burden and prefer AV shuttles as an additional way to travel.

In summary, the difference in people's attitudes toward AV shuttles from three perspectives may come from personal characteristics and transportation modes. On the one hand, respondents' judgment of their driving ability affects their attitudes as drivers. Respondents who have experienced the ability of AV technology to avoid collisions with people have a more positive attitude toward AV shuttles. Another aspect is the respondents' acceptance of AV shuttles as a new form of transportation. When respondents do not like shuttles as a traffic mode (participants with high income or who do not have to drive alone may not enjoy public transportation), they will be reluctant to act as riders but may be willing to try other autonomous technology.

### 4.4. Model estimation results of attitude change

The multinomial logit model is used to estimate participants' attitude change toward the comfort of the AV shuttle after a test ride, and the model estimation results are shown in Table 6. Seven variables are found to affect participants' attitude change on AV safety significantly. Therefore, the estimated utility functions are:

$$\begin{cases} V_0 = -1.323 X_0^{of} - 2.213 X_0^{hi} + 1.876 X_0^{nb} - 1.982 X_0^{ni} \\ V_1 = 1.990 - 1.381 X_1^{hi} - 1.741 X_1^{ni} \\ V_2 = 1.687 X_2^{lc} \end{cases} \quad (9)$$

The lack of constant in the no-change function establishes it as a 0 baseline. The negative constant in $V_0$ and $V_2$ shows people are more willing to keep their minds.

Although People with high incomes or long commute times have negative attitudes towards AV shuttles as riders, they are more likely to change their minds positively than others. The elasticity analysis shows that participants with an annual personal income higher than 100k are 54.95% more likely to change their minds positively. People with commute times longer than 30 minutes are 10.02% more likely to change their minds positively. This result shows that the AV shuttle operation companies might involve more people trying its service so that people who were originally optimistic about AV shuttle may change their attitude after the ride experience.

The number of abrupt brakes is an important factor that leads to negative change. A 1% increase in the number of abrupt brakes results in a 1.96% decrease in the probability of not changing the attitude. The abrupt brakes recorded in this study were sharp enough to be perceived by passengers. During the operation, the abrupt brakes are caused by several factors: 1) surrounding obstacles (e.g., overtaking vehicles, pedestrians on motor roads, grass or tree branches extending into the roads); 2) interrupted communication. Vehicles in these scenarios tend to adopt the behavior of slowing down sharply to zero and then starting to drive again. Even though all passengers were required to wear seat belts and were informed of the risk of sharp braking in advance, the abrupt brakes of the AV shuttle were unexpected and much more than common human-driven buses. This clearly had a significant influence on the rider's attitude of comfort.

Table 6 Multinomial Logit Estimation Results for Change of Attitude toward the comfort of the AV shuttle as riders (parameters defined for [NC] Negative change; [NO] Non-change function; [PC] Positive change).

| Variable Description | Estimated Parameter | t statistic |
|---|---|---|
| Negative change constant [NC] | -2.521 | -4.82 |
| Positive change constant [PC] | -0.912 | -3.26 |
| Variables that do not vary across alternate outcomes | | |
| High-income indicator (1 if the participant has annual personal income higher than 100k, 0 otherwise) [NC] | -1.907 | -3.14 |
| Number of abrupt brakes [NC] | 1.985 | 5.60 |
| High-income indicator (1 if the participant has annual personal income higher than 100k, 0 otherwise) [NO] | -0.987 | -2.34 |
| Long commute time indicator (1 if participant's commute time is longer than 30 minutes, 0 otherwise) [PC] | 1.614 | 3.29 |
| Number of observations | 161 | |
| Log-likelihood at convergence [LL(β)] | -138.372 | |

The multinomial logit model results of participants' attitude change toward comfort as pedestrians are shown in Table 7.

In the model of participants' attitude change toward comfort as pedestrians, people also prefer to keep their minds. Similar to the model of participants' attitude change toward the comfort of the AV shuttle, people with high incomes and long commute times are more seemingly to change their minds positively, and the number of abrupt brakes leads to negative change. Besides, people with automated riding experience are less likely to negatively change their attitude of comfort as pedestrians. This might be because they had preconceived notions about how the AV shuttle would function, and the test ride experience fulfilled their expectations.

Table 7 Multinomial Logit Estimation Results for Change of Attitude toward the comfort as Pedestrians (parameters defined for [NC] Negative change; [NO] Non-change function; [PC] Positive change).

| Variable Description | Estimated Parameter | t statistic |
|---|---|---|
| Negative change constant [NC] | -2.331 | -4.36 |
| Positive change constant [PC] | -0.931 | -2.07 |
| Variables that do not vary across alternate outcomes | | |
| High-income indicator (1 if the participant has annual personal income higher than 100k, 0 otherwise) [NC] | -1.388 | -2.07 |
| Number of abrupt brakes [NC] | 0.771 | 2.41 |
| Automated riding experience indicator (1 if the participant has automated riding experience before the test ride, 0 otherwise) [NC] | 2.100 | 2.91 |
| High-income indicator (1 if the participant has annual personal income higher than 100k, 0 otherwise) [NO] | -1.002 | -2.41 |
| Automated riding experience indicator (1 if the participant has automated riding experience before the test ride, 0 otherwise) [NO] | 1.451 | 2.41 |
| Long commute time indicator (1 if participant's commute time is longer than 30 minutes, 0 otherwise) [PC] | 1.356 | 2.76 |
| Number of observations | 152 | |
| Log-likelihood at convergence [LL($\beta$)] | -131.11 | |

Estimation results for change of attitude toward the comfort as drivers are presented in Table 8. Participants also like maintaining their thoughts from the driver's perspective. Similar to previous results, people with long commute times are more likely to change their minds positively, and the number of abrupt brakes leads to negative change. Moreover, females are less likely to change their minds negatively.

Table 8 Multinomial Logit Estimation Results for Change of Attitude toward the comfort as drivers (parameters defined for [NC] Negative change; [NO] Non-change function; [PC] Positive change).

| Variable Description | Estimated Parameter | t statistic |
|---|---|---|
| Negative change constant [NC] | -0.642 | -1.35 |
| Positive change constant [PC] | -0.112 | -0.41 |
| Variables that do not vary across alternate outcomes | | |
| Number of abrupt brakes [NC] | 0.635 | 2.17 |
| Female indicator (1 if the participant is female, 0 otherwise) [NC] | -0.831 | -1.82 |
| Highly interested in new technologies indicator (1 if the participant is highly interested in new technologies, 0 otherwise) [NO] | 0.708 | 1.93 |
| Long commute time indicator (1 if participant's commute time is longer than 30 minutes, 0 otherwise) [PC] | 1.385 | 2.74 |
| Number of observations | 135 | |
| Log-likelihood at convergence [LL($\beta$)] | -133.54 | |

Table 9 presents elasticity comparisons across the three perspectives. Among the six variables in the table that were significant across all age groups, there are some noteworthy differences. An instance is the elasticity of the number of abrupt brakes. The number of abrupt brakes has the greatest impact on the driver's attitude. The number of abrupt brakes raised the likelihood of negative change in attitude by 27.14% from the perspective of AV shuttle riders and 25.35% from the perspective of pedestrians, but 86.87% from the perspective of drivers. This difference may be explained by the operation process. There is ample empirical evidence during the survey showing AV shuttle might impede other vehicles. Most extremely, one rear-ends accident occurred during the survey right after abrupt brakes of the AV shuttle. Although no injury or death was caused, all four participants who experienced the accident negatively changed their attitudes in both rider perception and driver perception. Because of the small number of survey participants who experienced the accident, the accident is not included as an influencing factor in the models. In other situations, even if the abrupt brakes did not cause an accident, they showed negative effects besides uncomfortable. For instance, being overtaken by other vehicles is one of the major causes of abrupt brake. The AV shuttle is unable to accurately judge and predict the safety of this situation, so it will stop sharply and wait until the overtaking vehicle is far away before resuming driving. This kind of abrupt brake let the riders aware of the dissatisfaction of the following vehicles because of the low cruising speed. This shows that people are more worried about the disruption of traffic order and the instability of the vehicle embodied in the abrupt brakes than the discomfort caused by the abrupt brakes to passengers. In addition, high-income people are more likely to change their attitudes, especially as passengers. This may be because high-income people do not rely on shuttles for travel, and once the AV shuttles are not comfortable enough, they will choose other traffic modes.

Table 9 Comparison of elasticities for variables in the models of three different perspectives (elasticities are specific to [NC] Negative change; [NO] Non-change function; [PC] Positive change).

| From the Perspective of AV Shuttle riders | Elasticity | From the Perspective of Pedestrians | Elasticity | From the Perspective of Drivers | Elasticity |
|---|---|---|---|---|---|
| *Variables significant in all three models* | | | | | |
| Number of abrupt brakes [NC] | 27.14 | Number of abrupt brakes [NC] | 25.35 | Number of abrupt brakes [NC] | 86.87 |
| Long commute time indicator [PC] | 10.02 | Long commute time indicator [PC] | 9.21 | Long commute time indicator [PC] | 9.05 |
| *Variables significant in two models* | | | | | |
| High-income indicator [NC] | -47.39 | High-income indicator [NC] | -46.17 | | |
| High-income indicator [NO] | -183.9 | High-income indicator [NO] | -17.94 | | |
| *Variables significant in one model* | | | | | |
| | | Automated riding experience indicator [NC] | 11.41 | | |
| | | Automated riding experience indicator [NO] | 3.40 | | |
| | | | | Female indicator [NC] | 20.98 |
| | | | | Highly interested in new technologies indicator [NO] | 3.29 |

## 5. CONCLUSIONS AND FUTURE RESEARCH

Unlike private AVs that focus on tapping potential customers, AV shuttle is more concerned with the public trust from various groups in society. Understanding the factors that influence public attitudes toward AV shuttles is critical to increasing confidence in the technology and aiding its future deployment. This research collected information from 161 participants who experienced AV shuttle services, including personal characteristics, attitudes toward AV shuttles before a test ride, and attitudes toward AV shuttles after a test ride. Statistical models were used to analyze the factors influencing participants' initial attitudes and factors influencing attitude change. All significant variables found at least in one of the initial attitudes and attitude change models are summarized in Table 10. Our main findings are as follows:

1. Abrupt brakes of AV shuttle during the demonstration ride deteriorate the attitudes about the comfort of AV shuttles, especially from the driver's perspective.
2. Participants feel less comfortable sharing the rights of way with AV shuttles as drivers than taking the AV shuttle or sharing the rights of way with AV shuttles as pedestrians, for both the initial attitude and the change of attitude after taking the test ride of AV shuttles.
3. Participants with long commute times or high income tend to be conservative on AV shuttle comfort, but their attitudes significantly changed toward the positive side after a successful ride.
4. The seniors and females are more conservative about sharing the rights of way with AV shuttles, but successful rides also allow them to change their minds positively.

Table 10 Summary of initial attitude and attitude change models' findings.

| Variable Description | Initial Attitude of Comfort | | | Positive Change of the Comfort Attitude | | |
|---|---|---|---|---|---|---|
| | From the perspective of AV shuttle riders | From the perspective of pedestrians | From the perspective of drivers | From the perspective of AV shuttle riders | From the perspective of pedestrians | From the perspective of drivers |
| Senior indicator (1 if the participant is over age 50, 0 otherwise) | - | n | - | n | n | n |
| Female indicator (1 if the participant is female, 0 otherwise) | n | n | - | n | n | + |
| High-income indicator (1 if the participant has annual personal income higher than 200k, 0 otherwise) | - | n | n | + | + | n |
| Long commute time indicator (1 if participant's commute time is longer than 30 minutes, 0 otherwise) | - | - | - | + | + | + |
| Drive alone indicator (1 if the main commute mode of the participant is driving alone, 0 otherwise) | + | n | n | n | n | n |
| Automated riding experience indicator (1 if the participant has automated riding experience before the test ride, 0 otherwise) | n | + | n | n | - | n |
| Highly interested in new technologies indicator (1 if the participant is highly interested in new technologies, 0 otherwise) | + | + | + | n | n | - |
| Number of abrupt brakes | n | n | n | - | - | - |

Note: In the initial attitude model, "+" indicates the variable with a positive effect on the initial attitude, "-" with a negative effect on the initial attitude; In the attitude change model, "+" indicates the variable contributing to positive attitude change, "-" maintaining the same attitude or worse; "n" indicates no significant effects.

Our results have several implications on the policymaking of future AV technology:
1. Standardized tests on the operation quality of AV shuttles may be necessary before AV shuttles are permitted for public roads. Experimental results reveal that the poor operation comfort of AV shuttles significantly undermines people's attitudes toward AV shuttles. Therefore, AV testing standards established by policymakers could include the comfort evaluation of the rider. The comfort of riders could be evaluated using the human body vibration test (Cardinale & Pope, 2003). For example, the severity, frequency, and duration of AV riders' exposure to unsafe vibration could be evaluated and limited to avoid the negative effects of immature technology.
2. Vulnerable road users' acceptance of AV shuttles needed to be considered in relevant AV planning decisions. Experiments have shown that people perceive sharing the road with AV shuttles even more negatively than riding the AV shuttle. To address these concerns, AV manufacturers could consider human factors when designing AV shuttles and consider the behavior of other road users comprehensively. AV testing standards might not only focus on the competency of AV shuttles to successfully achieve driving tasks under every possible environment and traffic condition but also address whether AV shuttles' behavior has negative impacts on surrounding road users. For example, a slow operating speed may induce other drivers' lane-changing tendency, and sudden speed changes may deteriorate the trust of surrounding road users on AV shuttles. Existing AV testing studies often neglect the perspective of other road users (Nowakowski et al., 2015). Moreover, insurance companies may also consider safety when AV shuttles interact with other road users when insuring AV shuttles.
3. This work pinpoints groups with positive initial attitudes toward AV shuttles. This will help AV shuttle operating agencies identify targeted customers with positive initial attitudes toward AV shuttles and select pilot sites for AV shuttles. For example, our research shows participants with long commute times or high income are more likely to improve their attitude towards AV shuttle via test rides, which could help select AV shuttle demonstration sites; Our research also indicates that participants with high interest in new technologies have a more positive attitude towards AV shuttles. Thus, AV shuttle operators can place their advertisements near technology-related news, videos, or self-published content. In the future, the potential user targeting of AV shuttles can also help city planners to plan AV shuttle routes rationally and improve the utilization of AV shuttles.

Since significant differences can be observed in speed, operating hours, and fares between the studied AV shuttle and future regularly operated AV shuttles, future research can be conducted in several directions. First, AV shuttles operating in many cities are reported to have a low-speed limit (Kolodge et al., 2020). The over-slow operating speed may prevent AV shuttle from a practical solution to daily transportation and lead to more overtaking, thus further disrupting the traffic. Future research could assess how AV driving modes affect other vehicles and look into ways to lessen their negative effects while retaining the safety and comfort of AV shuttles. Second, the AV shuttles investigated in this study are not allowed to travel in the rain or snow. However, prior research (Ahmed et al., 2020) has demonstrated that equipment or systems failure in poor weather (storm, high wind, snow, rain, etc.) is a significant public concern. Transit agencies could focus on rigorous assessment and validation of AV shuttles in poor weather conditions to allay public fears. The next step of research might focus on generating extreme conditions for AV training to improve their stability in extreme weather. The third point is that existing AV demonstrations do not charge riders. This free-ride model is not sustainable, so the company will need to look at pricing reasonableness in the future to maximize revenue. The effect of AV fares on mode selection and transportation equity could be the subject of furture research.

## ACKNOWLEDGMENTS

The authors thank the National Science Foundation Grants CMMI #1932452 and the National Institute for Congestion Reduction #01751326 for funding this study.

Penmetsa, P., Adanu, E. K., Wood, D., Wang, T., & Jones, S. L. (2019). Perceptions and expectations of autonomous vehicles – A snapshot of vulnerable road user opinion. *Technological Forecasting and Social Change*, *143*(March), 9–13. https://doi.org/10.1016/j.techfore.2019.02.010

Pyrialakou, V. D., Gkartzonikas, C., Gatlin, J. D., & Gkritza, K. (2020). Perceptions of safety on a shared road: Driving, cycling, or walking near an autonomous vehicle. *Journal of Safety Research*, *72*, 249–258. https://doi.org/10.1016/j.jsr.2019.12.017

Rahman, M. T., & Dey, K. (2022). Perceptions of Vulnerable Roadway Users on Autonomous Vehicle Regulations. *SSRN Electronic Journal*. https://doi.org/10.2139/ssrn.4051556

Reig, S., Norman, S., Morales, C. G., Das, S., Steinfeld, A., & Forlizzi, J. (2018). A field study of pedestrians and autonomous vehicles. *Proceedings - 10th International ACM Conference on Automotive User Interfaces and Interactive Vehicular Applications, AutomotiveUI 2018*, 198–209. https://doi.org/10.1145/3239060.3239064

Salonen, A. O. (2018). Passenger's subjective traffic safety, in-vehicle security and emergency management in the driverless shuttle bus in Finland. *Transport Policy*, *61*(April 2017), 106–110. https://doi.org/10.1016/j.tranpol.2017.10.011

Sheela, P. V., & Mannering, F. (2020). The effect of information on changing opinions toward autonomous vehicle adoption: An exploratory analysis. *International Journal of Sustainable Transportation*, *14*(6), 475–487. https://doi.org/10.1080/15568318.2019.1573389

Shi, X., Wang, Z., Li, X., & Pei, M. (2020). *The Effect of Ride Experience on Changing Opinions Toward Autonomous Vehicle Safety*. *July 2021*, 1–10. https://doi.org/10.13140/RG.2.2.15902.56648/1

Templeton, B. (2020). *EasyMile Self-Driving Shuttle Banned After Sudden Stop Hurts Passenger — Are Seatbelts Needed?* https://www.forbes.com/sites/bradtempleton/2020/03/03/easymile-self-driving-shuttle-banned-after-sudden-stop-hurts-passenger---are-seatbelts-needed/?sh=2bcc6c677ad5

Washington, S., Karlaftis, M., Mannering, F., & Anastasopoulos, P. (2020). *Statistical and econometric methods for transportation data analysis*. Chapman and Hall/CRC.

Xing, Y., Zhou, H., Han, X., Zhang, M., & Lu, J. (2022). What influences vulnerable road users' perceptions of autonomous vehicles? A comparative analysis of the 2017 and 2019 Pittsburgh surveys. *Technological Forecasting and Social Change*, *176*(August 2020), 121454. https://doi.org/10.1016/j.techfore.2021.121454

Xu, Z., Zhang, K., Min, H., Wang, Z., Zhao, X., & Liu, P. (2018). What drives people to accept automated vehicles? Findings from a field experiment. *Transportation Research Part C: Emerging Technologies*, *95*(June), 320–334. https://doi.org/10.1016/j.trc.2018.07.024
19

# APPENDIX: Questionnaire

June, 2021 · Clearwater

## Survey on Autonomous Bus

This survey will take around 3 minutes to complete. Only aggregated results from the survey may be published without revealing any of your individual information. Thanks for your time.

### General information

1. Please select your age:

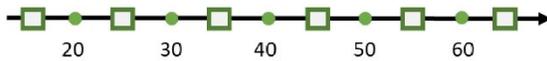

2. Please select your gender:
   ☐ Female  ☐ Male  ☐ Prefer to not answer

3. Please select your highest education:

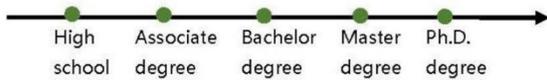

4. Please select your annual personal income:

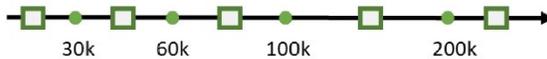

5. Please select your average commute time:
   ☐ shorter than 10 minutes
   ☐ 10 ~ 30 minutes
   ☐ 30minutes ~ 1 hour
   ☐ more than 1 hour

6. Please select your main commute mode:
   ☐ Drive alone      ☐ Rail transit
   ☐ Share ride       ☐ Walk
   ☐ Taxi             ☐ Bicycle
   ☐ Bus              ☐ Other

7. Have you ever had an autonomous riding experience? Such as autopilot?
   ☐ Yes              ☐ No

8. Are you interested in new technologies, such as robots?

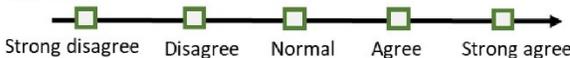

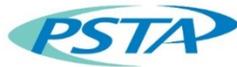

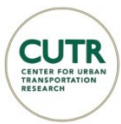 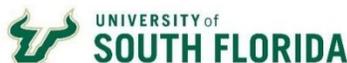

### Before your test ride

9. Your perception of comfort inside the autonomous bus (before test ride)

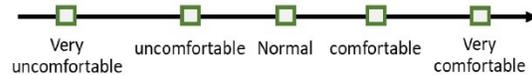

10. Your perception of comfort as a pedestrian crossing the road with autonomous buses (before test ride)

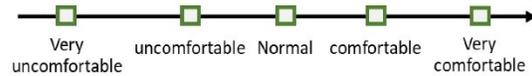

☐ N/A. No experience as a pedestrian

11. Your perception of comfort as a driver driving on the road with autonomous buses (before test ride)

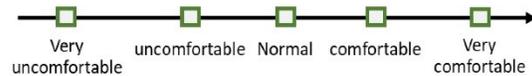

☐ N/A. No experience as a driver

### After your test ride

12. Your perception of comfort inside the autonomous bus (after test ride)

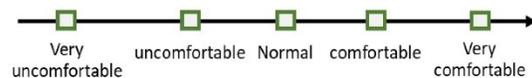

13. Your perception of comfort as a pedestrian crossing the road with autonomous buses (after test ride)

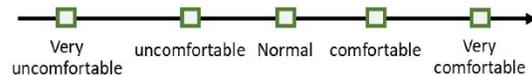

☐ N/A. No experience as a pedestrian

14. Your perception of comfort as a driver driving on the road with autonomous buses (after test ride)

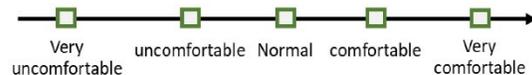

☐ N/A. No experience as a driver